\documentclass[11pt,preprint]{aastex}

\def\lsim{\lower.5ex\hbox{$\; \buildrel < \over \sim \;$}}
\def\gsim{\lower.5ex\hbox{$\; \buildrel > \over \sim \;$}}

\shorttitle{Delayed X- and Gamma-Ray Lines from Solar Flare Radioactivity}
\shortauthors{Tatischeff et al.}

\begin{document}


\title{Delayed X- and Gamma-Ray Line Emission \\ from Solar Flare Radioactivity}


\author{V. Tatischeff}
\affil{CSNSM, IN2P3-CNRS and Universit\'e Paris-Sud, F-91405 Orsay Cedex,
France}
\email{tatische@csnsm.in2p3.fr}

\author{B. Kozlovsky}
\affil{School of Physics and Astronomy, Tel Aviv University, Ramat Aviv, 
Tel Aviv 69978, Israel}

\author{J. Kiener}
\affil{CSNSM, IN2P3-CNRS and Universit\'e Paris-Sud, F-91405 Orsay Cedex,
France}

\and

\author{R. J. Murphy}
\affil{E. O. Hulburt Center for Space Research, Code 7650, 
Naval Research Laboratory, Washington, DC 20375 }


\begin{abstract}

We have studied the radioactive line emission expected from solar active regions
after large flares, following the production of long-lived radioisotopes by nuclear
interactions of flare-accelerated ions. This delayed X- and gamma-ray line 
emission can provide unique information on the accelerated particle composition 
and energy spectrum, as well as on mixing processes in the solar atmosphere. Total 
cross sections for the formation of the main radioisotopes by proton, $^3$He and
$\alpha$-particle reactions are evaluated from available data combined with
nuclear reaction theory. Thick-target radioisotope yields are provided in 
tabular form, which can be used to predict fluxes of all of the major delayed lines 
at any time after a gamma-ray flare. The brightest delayed line for days after the 
flare is found to be the 511~keV positron-electron annihilation line resulting from 
the decay of several $\beta^+$ radioisotopes. After $\sim$2~days however, the 
flux of the $e^+$--$e^-$ annihilation line can become lower than that of the 
846.8~keV line from the decay of $^{56}$Co into $^{56}$Fe. Our study has revealed 
other delayed gamma-ray lines that appear to be promising for detection, 
e.g. at 1434 keV from the radioactivity of both the isomer $^{52}$Mn$^m$ 
($T_{1/2}$=21.1~min) and the ground state $^{52}$Mn$^g$ ($T_{1/2}$=5.59~days), 
1332 and 1792~keV from $^{60}$Cu ($T_{1/2}$=23.7~min), and 931.1~keV from 
$^{55}$Co ($T_{1/2}$=17.5~hours). The strongest delayed X-ray line is found to be 
the Co K$\alpha$ at 6.92~keV, which is produced from both the decay of the isomer 
$^{58}$Co$^m$ ($T_{1/2}$=9.04~hours) by the conversion of a K-shell electron and 
the decay of $^{57}$Ni ($T_{1/2}$=35.6~hours) by orbital electron capture.
Prospects for observation of these lines with {\it RHESSI} or future space 
instruments are discussed. 
\end{abstract}


\keywords{nuclear reactions, nucleosynthesis, abundances -- Sun: flares -- 
Sun: X-rays, gamma rays}


\section{Introduction}

Gamma-ray lines from solar flares were first observed in 1972 with the 
gamma-ray spectrometer (GRS) aboard the {\it OSO-7} satellite \citep{chu73}.
Since then, repeated observations with various space missions, including the
{\it Solar Maximum Mission}/GRS (e.g. Share \& Murphy 1995), all four {\it
Compton Gamma Ray Observatory} instruments (e.g. Share, Murphy, \& Ryan 1997)
and the {\it Reuven Ramaty High Energy Solar Spectroscopic Imager }({\it
RHESSI}; e.g. Lin et al. 2003), have firmly established gamma-ray astronomy 
as an important tool for studying the active sun. Prompt gamma-ray lines are 
produced from deexcitation of nuclei excited by nuclear interactions of 
flare-accelerated particles with the solar atmosphere. Detailed spectroscopic 
analyses of this emission have furnished valuable information on the composition
of the ambient flare plasma, as well as on the composition, energy spectrum and 
angular distribution of the accelerated ions (e.g. Ramaty \& Mandzhavidze 2000; 
Share \& Murphy 2001; Lin et al. 2003; Kiener et al. 2006). Additional
information about the density and temperature of the ambient plasma is
obtained from the positron-electron annihilation line at 0.511 MeV (Murphy et 
al. 2005) and the neutron capture line at 2.223 MeV (Hua et al. 2002 and
references therein). 

The bombardment of the solar atmosphere by flare-accelerated ions can also
synthesize radioactive nuclei, whose decay can produce observable, delayed 
gamma-ray lines in the aftermath of large flares. One of the most promising of 
such lines is at 846.8~keV resulting from the decay of $^{56}$Co (half-life 
$T_{1/2}$=77.2~days) into the first excited state of $^{56}$Fe \citep{ram00,koz02}. 
\citet{ram00} calculated the time dependence of the 846.8~keV line emission that 
would have been expected after the 6 X-class flares of June 1991. Smith et al. are 
now searching for this delayed line emission with the {\it RHESSI} 
spectrometer after the very intense series of flares that occured between 2003 
October 28 and November 4 (the analysis is in progress, D. Smith 2006, private
communication). 

The observation of solar radioactivity can be important for at 
least two reasons. First, the radioisotopes can serve as tracers to study
mixing processes in the solar atmosphere \citep{ram00}. Additionally, 
their detection should provide a new insight into the spectrum and fluence of
flare-accelerated ions. In particular, since the accelerated heavy nuclei
are believed to be significantly enhanced as compared to the ambient medium 
composition (e.g. Murphy et al. 1991), the radioisotopes are expected to be 
predominantly produced by interactions of fast heavy ions with ambient
hydrogen and helium. Thus, the delayed line emission can provide a valuable
measurement of the accelerated metal enrichment. 

We performed a systematic study of the radioactive line emission expected
after large solar flares. In addition to gamma-ray lines emitted from 
deexcitation of daughter nuclei, we considered radioactivity X-ray lines 
that can be produced from the decay of proton-rich isotopes by orbital electron 
capture or the decay of isomeric nuclear levels by emission of a conversion 
electron. We also treated the positron-electron annihilation line resulting from 
the decay of long-lived $\beta^+$-emitters. The radioisotopes which we studied are 
listed in Table~1, together with their main decay lines. We selected
radioactive X- and gamma-ray line emitters that can be significantly produced in
solar flares (see \S~2) and with half-lives between $\sim$10~min, which 
is the typical duration of large gamma-ray flares \citep{ves99}, and 77.2~days 
($^{56}$Co half-life). We neglected radioisotopes with mean lifetime $\tau_r$ 
greater than that of $^{56}$Co, because (1) their activity 
($\dot{N_r}=N_r/\tau_r$) is lower and (2) their chance of surviving at the solar 
surface is also lower.

In \S~2, we present the total cross sections for the production of the most
important radioactive nuclei. In \S~3, we describe our thick-target yield 
calculations of the radioisotope synthesis. The results for the
delayed line emission are presented in \S~4. Prospects for observations are 
discussed in \S~5.

\section{Radioisotope production cross sections}

All of the radioactive nuclei shown in Table~1 can be significantly produced in 
solar flares by H and He interactions with elements among the most abundant of the 
solar atmosphere and accelerated particles: He, C, N, O, Ne, Mg, Al, Si, S, Ar, Ca, 
Cr, Mn, Fe and Ni\footnote{\citet{kuz05} recently claimed that nuclear interactions
between fast and ambient heavy nuclei can be important for the formation 
of rare isotopes in solar flares. We evaluated the significance of these 
reactions by using the universal parameterization of total reaction cross sections 
given by \citet{tri96}. Assuming a thick target interaction model, a power-law
source spectrum for the fast ions and standard compositions for the
ambient and accelerated nuclei (see \S~3), we found that the heavy ion
collisions should contribute less than a few percent of the total
radioisotope production and can therefore be safely neglected.}. 
We did not consider the production of radioisotopes with atomic 
number $Z>30$. We also neglected a number of very neutron-rich nuclei (e.g. 
$^{28}$Mg, $^{38}$S...), whose production in solar flares should be very low.

Most of the radioisotopes listed in Table~1 are proton-rich, positron emitters.
Their production by proton and $\alpha$-particle reactions with the abundant
constituents of cosmic matter was treated in detail by Kozlovsky, Lingenfelter, 
\& Ramaty (1987). The work of these authors was extended by Kozlovsky, 
Murphy, \& Share (2004) to include $\beta^+$-emitter production from the most 
important reactions induced by accelerated $^3$He. However, new laboratory 
measurements have allowed us to significantly improve the evaluation of a
number of cross sections for the production of important positron
emitters. In the following, we present updated cross sections for the
formation of $^{34}$Cl$^m$, $^{52}$Mn$^g$, $^{52}$Mn$^m$, $^{55}$Co,
$^{56}$Co, $^{57}$Ni, $^{58}$Co$^g$, $^{60}$Cu, and $^{61}$Cu, by proton, 
$^3$He and $\alpha$ reactions in the energy range 1--10$^3$ MeV nucleon$^{-1}$. 
In addition, we evaluate cross sections for the production of the 4 
radioactive nuclei of Table~1 which are not positron emitters: $^7$Be, $^{24}$Na, 
$^{56}$Mn and $^{58}$Co$^m$. 

The reactions which we studied are listed in Table~2. We considered proton, 
$^3$He and $\alpha$-particle interactions with elements of atomic numbers close 
to that of the radioisotope of interest and that are among the most abundant of 
the solar atmosphere. Spallation reactions with more than 4 outgoing 
particles were generally not selected, because their cross sections are usually 
too low and their threshold energies too high to be important for solar flares. 
We generally considered reactions with elements of natural isotopic compositions, 
because, except for H and the noble gases, the terrestrial isotopic 
compositions are representative of the solar isotopic abundances \citep{lod03}. 
Furthermore, most of the laboratory measurements we used were performed 
with natural targets. 

Most of the cross section data were extracted from the EXFOR database for
experimental reaction data\footnote{See http://www.nndc.bnl.gov/exfor/.}. When
laboratory measurements were not available or did not cover the full energy
range, we used 3 different nuclear reaction codes to obtain theoretical 
estimates. Below a few hundred MeV (total kinetic energy of the fast particles), 
we performed calculations with both EMPIRE-II (version 2.19; Herman et al. 
2004) and TALYS (version 0.64; Koning, Hilaire, \& Duijvestijn 2005). These 
computer programs account for major nuclear reaction models for direct,
compound, pre-equilibrium and fission reactions. They include comprehensive 
libraries of nuclear structure parameters, such as masses, discrete level
properties, resonances and gamma-ray parameters. The TALYS and EMPIRE-II 
calculations were systematically compared with available data and the agreement 
was generally found to be better than a factor of 2. We obtained, however, more 
accurate predictions for isomeric cross sections with TALYS than with 
EMPIRE-II. 

Above the energy range covered by TALYS and EMPIRE-II, we used the "Silberberg 
\& Tsao code" (Silberberg, Tsao, \& Barghouty 1998 and references therein) when 
experimental cross section data for proton-nucleus spallation reactions were 
lacking. This code is based on the semiempirical formulation originally developed 
by \citet{sil73} for estimates of cross sections needed in cosmic-ray physics. It 
has been updated several times as new cross sections measurements have become 
available \citep{sil98}. For spallation reactions induced by 
$\alpha$-particles above $\sim$100~MeV nucleon$^{-1}$, we used the approximation 
(Silberberg \& Tsao 1973)
\begin{equation}
\sigma_\alpha(E)=X\sigma_p(4E)~,
\end{equation}
where $E$ is the projectile kinetic energy per nucleon, $\sigma_\alpha$ and 
$\sigma_p$ are the cross sections for the $\alpha$-particle- and 
proton-induced reactions leading to the same product, and
\begin{equation}
X = \left\{ \begin{array}{ll}
1.6 & \rm{for~} \Delta A \lsim 3 \\
2 & \rm{for~} \Delta A > 3~,
\end{array} \right.
\end{equation}
where $\Delta A$ is the difference between target and product masses. 

\subsection{$^7$Be Production}

The relevant cross sections for $^7$Be production are shown in Figure~1. 
The cross section for the reaction $^4$He($\alpha$,$n$)$^7$Be (dashed curve 
labeled "$^4$He" in Fig.~1) is from the measurements of \citet{kin77} from 9.85 
to 11.85~MeV nucleon$^{-1}$ and Mercer et al. (2001 and references therein) above 
$\sim$15.4~MeV nucleon$^{-1}$. The cross sections for the proton reactions with 
$^{12}$C, $^{14}$N and $^{16}$O (solid curves in Fig.~1) are from the extensive
measurements of Michel et al. (1997 and references therein). The cross sections
for the $\alpha$-particle reactions with $^{12}$C, $^{14}$N and $^{16}$O 
(dashed curves labeled "$^{12}$C", "$^{14}$N" and "$^{16}$O" in Fig.~1) are 
from the measurements of \citet{lan95} below 42 MeV nucleon$^{-1}$ and from 
TALYS calculations at 50 and 62.5~MeV nucleon$^{-1}$. At higher energies, we 
used the data compilation of \citet{rea84} and assumed the $^7$Be production 
cross sections to be half of the isobaric cross sections for producing the mass 
$A$=7 fragment from spallation of the same target isotope. The cross section 
for the reaction $^{12}$C($^3$He,$x$)$^7$Be (dotted curve in Fig.~1) is from 
\citet{dit94} below 9.1~MeV nucleon$^{-1}$ and from TALYS calculations from 
10 to 83.3~MeV nucleon$^{-1}$. At higher energies, we extrapolated the cross
section assuming the same energy dependence as the one for the 
$^{12}$C($\alpha$,$x$)$^7$Be reaction. We neglected the production of $^7$Be from 
$^3$He reactions with $^{14}$N and $^{16}$O. 

\subsection{$^{24}$Na Production}

The relevant cross sections for $^{24}$Na production are shown in Figure~2. The 
cross section for the reaction $^{25}$Mg($p$,2$p$)$^{24}$Na is from \citet{mea51} 
below 105 MeV, \citet{ree69} in the energy range 105--300~MeV and above 400~MeV, 
and \citet{kor70} at 300 and 400~MeV. The cross section for the reaction 
$^{26}$Mg($p$,2$pn$)$^{24}$Na is also from \citet{mea51} and \citet{kor70} below
400~MeV. Its extrapolation at higher energies was estimated from calculations
with the Silberberg \& Tsao code. The cross sections for the proton reactions 
with $^{27}$Al and $^{nat}$Si are from the measurements of Michel et al. (1997).
The cross section for the reaction $^{nat}$Mg($\alpha$,$x$)$^{24}$Na is from
the data of \citet{lan95} below 42 MeV nucleon$^{-1}$ and from TALYS calculations 
at 50 and 62.5~MeV nucleon$^{-1}$. Above 100 MeV nucleon$^{-1}$, the 
$\alpha$+$^{nat}$Mg cross section was estimated from equations~(1) and (2), and 
the $p$+$^{25}$Mg and $p$+$^{26}$Mg cross sections discussed above. The cross 
sections for the reactions $^{22}$Ne($\alpha$,$pn$)$^{24}$Na and 
$^{22}$Ne($^3$He,$p$)$^{24}$Na are not available in the literature and were 
estimated from calculations with the TALYS and EMPIRE-II codes, respectively. 

\subsection{$^{34}$Cl$^m$ Production}

Shown in Figure~3 are cross sections for production of the first excited
(isomeric) state of $^{34}$Cl ($^{34}$Cl$^m$, $T_{1/2}=32$~min) at an 
excitation energy of 146.4~keV. The cross section data for these reactions are 
scarce. The cross section for the reaction $^{32}$S($^3$He,$p$)$^{34}$Cl$^m$ was 
measured by \citet{lee74} from 1.4 to $\sim$7.3~MeV nucleon$^{-1}$. Its rapid fall 
at higher energies was estimated from TALYS calculations. The cross section 
of the reaction $^{34}$S($p$,$n$)$^{34}$Cl$^m$ was measured by \citet{hin52} from 
threshold to $\sim$90~MeV. However, as the decay scheme of $^{34}$Cl was not well 
known in 1952, it is not clear which fraction of the isomeric state was populated 
in this experiment. We thus did not use these data, but have estimated the cross
section from TALYS calculations. We also used theoretical evaluations from
the TALYS and the Silberberg \& Tsao codes for the reactions 
$^{32}$S($\alpha$,$pn$)$^{34}$Cl$^m$ and $^{sol}$Ar($p$,$x$)$^{34}$Cl$^m$. In 
this latter reaction, the notation $^{sol}$Ar means Ar of solar isotopic 
composition\footnote{The solar isotopic composition of Ar and the other noble 
gases are very different from their terrestrial isotopic compositions (see 
Lodders 2003 and references therein).} and the cross section was obtained by 
weighting the cross sections for proton reactions with $^{36}$Ar, $^{38}$Ar 
and $^{40}$Ar by the relative abundances of these three isotopes in the solar 
system. 

\subsection{$^{52}$Mn$^{g,m}$ Production}

The production of the ground state of $^{52}$Mn ($^{52}$Mn$^g$, 
$T_{1/2}=5.59$~days) and of the isomeric level at 377.7~keV ($^{52}$Mn$^m$, 
$T_{1/2}=21.1$~min) are both important for the delayed line emission of solar 
flares. The relevant cross sections are shown in Figure~4. The data for the 
production of the isomeric pair $^{52}$Mn$^g$ and $^{52}$Mn$^m$ in
$p$+$^{nat}$Cr collisions are from \citet{win62} from 5.8 MeV to 10.5~MeV; 
\citet{wes87} from 6.3 to $\sim$26.9~MeV; \citet{kle00} from $\sim$17.4 to 
$\sim$38.1~MeV; and \citet{reu69} at 400~MeV. It is noteworthy that TALYS 
simulations for these reactions were found to be in very good agreement with 
the data, which demonstrates the ability of this code to predict accurate
isomeric state populations. The cross section for the reaction 
$^{nat}$Fe($p$,$x$)$^{52}$Mn$^g$ is from \citet{mic97}. We estimated the 
cross section for the production of the isomer $^{52}$Mn$^m$ in 
$p$+$^{nat}$Fe collisions by multiplying the cross section for the ground 
state production by the isomeric cross section ratio $\sigma_m / \sigma_g$ 
calculated with the TALYS code. The cross sections for the production of 
$^{52}$Mn$^g$ and $^{52}$Mn$^m$ from $\alpha$+$^{nat}$Fe interactions are also 
from TALYS calculations below 62.5~MeV nucleon$^{-1}$. They were extrapolated at 
higher energies using equations~(1) and (2), and the $p$+$^{nat}$Fe cross sections 
discussed above. Also shown in Figure~4 is the cross section for the reaction 
$^{nat}$Cr($^3$He,$x$)$^{52}$Mn$^m$, which is based on the data of 
\citet{fes94} below $\sim$11.7~MeV nucleon$^{-1}$ and TALYS calculations at higher
energies. 

\subsection{$^{56}$Mn Production}

The relevant cross sections for $^{56}$Mn production are shown in Figure~5. The
laboratory measurements for the production of this radioisotope are few. We used 
the experimental works of the following authors: \citet{wat79} for the reaction 
$^{55}$Mn($^3$He,2$p$)$^{56}$Mn from $\sim$3.8 to $\sim$12.9~MeV nucleon$^{-1}$;
Michel, Brinkmann, \& St\"uck (1983a) for the reaction 
$^{55}$Mn($\alpha$,2$pn$)$^{56}$Mn from 6.1 to $\sim$42.8~MeV nucleon$^{-1}$; and 
Michel, Brinkmann, \& St\"uck (1983b) for the reaction 
$^{nat}$Fe($\alpha$,$x$)$^{56}$Mn from $\sim$13.8 to $\sim$42.8~MeV nucleon$^{-1}$. 
The excitation functions for these 3 reactions were completed by
theoretical estimates from TALYS. The cross section for the reaction 
$^{57}$Fe($p$,2$p$)$^{56}$Mn is entirely based on nuclear model calculations, from 
EMPIRE-II below 100 MeV and the Silberberg \& Tsao code at higher energies. 

\subsection{$^{55}$Co, $^{56}$Co and $^{57}$Ni Production}

The relevant cross sections for production of $^{55}$Co, $^{56}$Co and $^{57}$Ni 
are shown in Figures~6, 7 and 8, respectively. The cross sections for the proton 
reactions with $^{nat}$Fe and $^{nat}$Ni are based on the data of \citet{mic97}. 
For $^{56}$Co production by $p$+$^{nat}$Fe and $p$+$^{nat}$Ni collisions, we also 
used the works of \citet{tak94} and \citet{tar91}, respectively. The cross sections
for the $\alpha$-particle reactions with $^{nat}$Fe are from \citet{tar03b} below 
10.75~MeV nucleon$^{-1}$, \citet{mic83b} in the energy range 
$\sim$12.3--42.8~MeV nucleon$^{-1}$ and TALYS calculations at 50 and 62.5~MeV 
nucleon$^{-1}$. These cross sections were extrapolated at higher energies
assuming that they have energy dependences similar to those of the $p$+$^{nat}$Fe 
reactions (see eqs.~[1] and [2]). The cross sections for the $\alpha$-particle 
reactions with $^{nat}$Ni are based on the data of \citet{mic83b}. For the reaction 
$^{nat}$Ni($\alpha$,$x$)$^{57}$Ni, we also used the measurements of
Tak\'acs, T\'ark\'anyi, \& Kovacs (1996) below $\sim$6.1~MeV nucleon$^{-1}$. The
procedure to estimate the $\alpha$+$^{nat}$Ni cross sections above 50~MeV 
nucleon$^{-1}$ was identical to the one discussed above for the $\alpha$+$^{nat}$Fe 
cross sections. The cross sections for the $^3$He reactions with $^{nat}$Fe are
based on the data of \cite{tar03a} from $\sim$4.1 to $\sim$8.5~MeV nucleon$^{-1}$, 
the data of \citet{haz65} from 1.9 to $\sim$19.8~MeV nucleon$^{-1}$, and
TALYS calculations. The measurements of \citet{haz65} were performed with 
targets enriched in $^{56}$Fe. To estimate the cross section for $^3$He+$^{nat}$Fe 
collisions from their data, we multiplied the measured cross section by 
0.92, the relative abundance of $^{56}$Fe in natural iron. We neglected the 
production of $^{55}$Co and $^{56}$Co by $^3$He+$^{nat}$Ni interactions. For the 
reaction $^{nat}$Ni($^3$He,$x$)$^{57}$Ni, we used the data of \citet{tak95} below 
$\sim$11.7~MeV nucleon$^{-1}$ and EMPIRE-II calculations to 80~MeV nucleon$^{-1}$. 
At higher energies, the cross section was extrapolated assuming an energy 
dependence similar to the one of the $^{nat}$Ni($\alpha$,$x$)$^{57}$Ni cross 
section.

\subsection{$^{58}$Co$^{g,m}$ Production}

The relevant cross sections for the production of the isomeric pair 
$^{58}$Co$^{g,m}$ are shown in Figures~9a and b. The isomeric state of $^{58}$Co
($^{58}$Co$^m$, $T_{1/2}=9.04$~hours) is the first excited level at 24.9~keV. It 
decays to the ground state by the conversion of a K-shell electron, thus producing 
a Co K$\alpha$ line emission at 6.92~keV. The ground state $^{58}$Co$^g$ having a 
much longer lifetime, $T_{1/2}=70.9$~days, we considered for its production the 
total cross section for the formation of the isomeric pair, 
$\sigma_t = \sigma_m + \sigma_g$. Isomeric cross section ratios 
$\sigma_m / \sigma_t$ were measured for various reaction channels by Sud\'ar \& 
Qaim (1996 and references therein). We used their data for the reaction 
$^{55}$Mn($\alpha$,$n$)$^{58}$Co$^{g,m}$ below $\sim$6.3~MeV nucleon$^{-1}$. At 
higher energies, we used the $\sigma_m$ and $\sigma_t$ measurements of 
\citet{mat65} and \citet{riz89}, respectively. The total cross section for the 
reaction $^{nat}$Fe($\alpha$,$x$)$^{58}$Co is based on the data of \citet{iwa62} 
from 4.4 to 9.65~MeV nucleon$^{-1}$ and \citet{mic83b} in the energy range 
$\sim$6.5--42.8~MeV nucleon$^{-1}$. In the absence of data for the isomer 
formation in $\alpha$+$^{nat}$Fe collisions, we estimated the cross 
section by multiplying the cross section for the total production of 
$^{58}$Co by the isomeric ratio $\sigma_m / \sigma_t$ calculated with the TALYS 
code. The total cross section for the reaction $^{nat}$Fe($^3$He,$x$)$^{58}$Co is 
from \citet{haz65} and \citet{tar03a} below $\sim$8.5~MeV nucleon$^{-1}$. At 
higher energies, it is based on TALYS calculations. The cross section for the 
isomeric state population in $^3$He+$^{nat}$Fe collisions is also from
simulations with the TALYS code. 

The cross sections for $^{58}$Co$^{g,m}$ production from proton, $^3$He and 
$\alpha$ reactions with $^{nat}$Ni are shown in Figure~9b. The total cross
sections $\sigma_t$ are based on the data of \citet{mic97}, \citet{tak95} below  
$\sim$11.7~MeV nucleon$^{-1}$, and \citet{mic83b} below 
$\sim$42.7~MeV nucleon$^{-1}$, for the proton, $^3$He and $\alpha$ reactions,
respectively. To extrapolate the cross sections for the $^3$He and 
$\alpha$ reactions, we used the TALYS code and the approximation described by
equations (1) and (2). Data for the isomeric state population are lacking and
we estimated the $\sigma_m$ cross sections as above, i.e. from TALYS
calculations of the isomeric ratios $\sigma_m / \sigma_t$. 

\subsection{$^{60}$Cu and $^{61}$Cu Production}

The relevant cross sections for the production of $^{60}$Cu and $^{61}$Cu are 
shown in Figure~10. The cross sections for the production of $^{60}$Cu are based 
on the data of \citet{bar75} below $\sim$17~MeV, \citet{mur78} below $\sim$8.8~MeV 
nucleon$^{-1}$, and \citet{tak95} below $\sim$11.7~MeV nucleon$^{-1}$, for the 
proton, $\alpha$-particle and $^3$He reactions with $^{nat}$Ni, respectively. The 
cross section for the reaction $^{nat}$Ni($^3$He,$x$)$^{61}$Cu is also from 
\citet{tak95} below $\sim$11.7~MeV nucleon$^{-1}$. The cross section for the 
reaction $^{nat}$Ni($\alpha$,$x$)$^{61}$Cu was constructed from the data of 
\citet{tak96} below $\sim$6.1~MeV nucleon$^{-1}$, \citet{mur78} from $\sim$2.5
to $\sim$9.2~MeV nucleon$^{-1}$, and \citet{mic83b} in the energy range
$\sim$4.2--42.7~MeV nucleon$^{-1}$. All these cross sections were extrapolated
to higher energies by the means of TALYS calculations.

\section{Radioisotope production yields}

We calculated the production of radioactive nuclei in solar flares assuming a
thick target interaction model, in which accelerated particles with given energy
spectra and composition produce nuclear reactions as they slow down in the solar 
atmosphere. Taking into account the nuclear destruction and catastrophic energy 
loss (e.g. interaction involving pion production) of the fast particles in the 
interaction region, the production yield of a given radioisotope $r$ can be 
written as (e.g. Parizot \& Lehoucq 1999): 
\begin{equation}
Q_r = \sum_{ij} n_j \int_0^{\infty} {dE v_i(E) \sigma_{ij}^r(E) \over \dot{E}_i(E)}
\int_E^{\infty} dE' N_i(E') \exp 
\bigg[- \int_E^{E'} {dE'' \over \dot{E}_i(E'') \tau_i^{ine}(E'')}\bigg]~,
\end{equation}
where $i$ and $j$ range over the accelerated and ambient particle species that 
contribute to the synthesis of the radioisotope considered, $n_j$ is the density 
of the ambient constituent $j$, $v_i$ is the velocity of the fast ion $i$, 
$\sigma_{ij}^r$ is the cross section for the nuclear reaction $j$($i$,$x$)$r$,
$\dot{E}_i$ is the energy loss rate for the accelerated particles of type $i$ in 
the ambient medium, $N_i$ is the source energy spectrum for these particles, and 
$\tau_i^{ine}$ is the energy dependent average lifetime of the fast ions of type
$i$ before they suffer inelastic nuclear collisions in the interaction region. 
As H and He are by far the most abundant constituents of the solar atmosphere,
we have
\begin{equation}
\tau_i^{ine} \cong {1 \over v_i(n_H\sigma_{iH}^{ine} + n_{He}\sigma_{iHe}^{ine})}~,
\end{equation}
where $\sigma_{iH}^{ine}$ and $\sigma_{iHe}^{ine}$ are the total inelastic cross
sections for particle $i$ in H and He, respectively. We used the cross sections
given by \citet{mos02} for the $p$--H and $p$--He total inelastic reactions and 
the universal parameterization of Tripathi et al. (1996,1999) for the other fast 
ions. 

The energy loss rate was obtained from
\begin{equation}
\dot{E}_i = v_i {Z_i^2(\rm{eff}) \over A_i} 
\bigg[ n_H m_H \bigg({dE \over dx}\bigg)_{pH} + 
n_{He} m_{He} \bigg({dE \over dx}\bigg)_{pHe} \bigg]~,
\end{equation}
where $(dE/dx)_{pH}$ and $(dE/dx)_{pHe}$ are the proton stopping powers (in
units of MeV g$^{-1}$ cm$^{2}$) in ambient H and He, respectively \citep{ber05}, 
$m_H$ and $m_{He}$ are the H- and He-atom masses, 
$Z_i(\rm{eff}) = Z_i[1-\exp(-137\beta_i/Z_i^{2/3})]$ is the equilibrium effective 
charge \citep{pie68}, $\beta_i=v_i/c$ is the particle velocity relative to that
of light, and $Z_i$ and $A_i$ are the nuclear charge and mass for particle
species $i$, respectively. Inserting equations (4) and (5) into equation (3), we
see that under the assumption of thick target interactions, the yields 
do not depend on the ambient medium density, but only on the
relative abundances $n_j/n_H$. We used for the ambient medium composition the
same abundances as Kozlovsky et al. (2004, Table~2). 

We took for the source energy spectrum of the fast ions an unbroken power law 
extending from the threshold energies of the various nuclear reactions 
up to $E_{max}$=1~GeV nucleon$^{-1}$: 
\begin{equation}
N_i(E)=C_i E^{-s} H(E_{max}-E)~,
\end{equation}
where the function $H(E)$ denotes the Heaviside step function and $C_i$ is the 
abundance of the accelerated particles of type $i$. We assumed the following 
impulsive-flare composition for the accelerated ions: we used for the abundances
of fast C and heavier elements relative to $\alpha$-particles the average 
composition of solar energetic particles (SEP) measured in impulsive events from 
interplanetary space (Reames 1999, Table 9.1), but we took the accelerated 
$\alpha/p$ abundance ratio to be 0.5, which is at the maximum of the range 
observed in impulsive SEP events. The choice of such a large $\alpha/p$ ratio is 
motivated by analyses of gamma-ray flares \citep{sha97,man97,man99}, showing a 
relatively strong emission in the line complex at $\sim$450~keV from 
$\alpha$-particle interactions with ambient $^4$He. The expected modifications of 
our results for higher proton abundances relative to $\alpha$-particles and 
heavier ions are discussed in \S~4. We performed calculations with an accelerated 
$^3$He/$\alpha$ abundance ratio of 0.5, which is typical of the accelerated 
$^3$He enrichment found in impulsive SEP events \citep{rea94,rea99}, as well as 
in gamma-ray flares (Share \& Murphy 1998; Manzhavidze et al. 1999). The 
resulting accelerated-particle composition is similar to the one used by 
\citet{koz04}, but slightly less enriched in heavy elements (e.g. Fe and Ni 
abundances are lower than those of these authors by 13\% and 29\%, respectively). 
The enhancement of the fast heavy elements is however still large relative to 
the ambient material composition. We have, for example, 
$C_{Fe}/C_p=137n_{Fe}/n_H$. 
 
Thick-target radioisotope yields are given in Table~3 for $s$=3.5, 2 and 5
(eq.~[6]). The first value is close to the mean of spectral index distribution as 
measured from analyses of gamma-ray line ratios \citep{ram96}, whereas the two 
other values are extreme cases to illustrate the dependence of the radioisotope 
production on the spectral hardness. The calculations were normalized to unit 
incident number of protons of energy greater than 5~MeV. For comparison, the 
last two lines of this table give thick-target yields for the production of the 
4.44 and 6.13 MeV deexcitation lines from ambient $^{12}$C and $^{16}$O, 
respectively. These prompt narrow lines are produced in reactions of energetic 
protons and $\alpha$-particles with ambient $^{12}$C, $^{14}$N, $^{16}$O and 
$^{20}$Ne (see Kozlovsky et al. 2002). We can see that, relative to these two 
gamma-ray lines, the production of most of the radioisotopes increases as the 
accelerated particle spectrum becomes harder (i.e. with decreasing $s$). This 
is because the radioactive nuclei are produced by spallation reactions at 
higher energies, on average, than the $^{12}$C and $^{16}$O line emission, 
which partly results from inelastic scattering reactions. 

Because of the enhanced heavy accelerated particle composition, most of the 
yield is from interactions of heavy accelerated particles with ambient H and 
He. For example, the contribution of fast Fe and Ni collisions with 
ambient H and He accounts for more than 90\% of the total $^{56}$Co production,
whatever the spectral index $s$.

Because we are interested in emission after the end of the gamma-ray flare, we 
show in the fifth column of Table~3 a factor $f_d$ which should be multiplied 
with the given yields to take into account the decay of the radioactive nuclei 
occuring before the end of the flare. It was calculated from the simplifying 
assumption that the radioisotope production rate is constant with time during 
the flare for a time period $\Delta t$. We then have
\begin{equation}
f_d={\tau_r \over \Delta t} (1-e^{-\Delta t / \tau_r})~,
\end{equation}
where $\tau_r$ is the mean lifetime of radioisotope $r$. In Table~3, $f_d$ is 
given for $\Delta t$=10~min. 

\section{Delayed X- and gamma-ray line emission}

Calculated fluxes of the most intense delayed lines are shown in Tables~4--6 for 
three different times after a large gamma-ray flare. The lines are given in
decreasing order of their flux for $s$=3.5.
The calculations were normalized to a total fluence of the summed 4.44 and 
6.13~MeV prompt narrow lines $\mathcal{F}_{4.4+6.1}$=300 photons~cm$^{-2}$, 
which is the approximate fluence observed in the 2003 October 28 flare with 
{\it INTEGRAL}/SPI \citep{kie06}. The flux of a given delayed line $l$ produced by 
the decay of a radioisotope $r$ at time $t$ after the end of the nucleosynthesis 
phase was obtained from
\begin{equation}
F_l(t)={\mathcal{F}_{4.4+6.1} Q_r f_d I_l^r \over Q_{4.4+6.1} \tau_r} 
e^{-t / \tau_r}~,
\end{equation}
where $Q_r$ and $Q_{4.4+6.1}$ are the yields (Table~3) of the parent 
radioisotope and summed prompt $^{12}$C and $^{16}$O lines, respectively, 
and $I_l^r$ is the line branching ratio (the percentages shown in Table~1). The 
factor $f_d$ was calculated for a flare duration of 10~min (Table~3). The
calculated fluxes do not take into account attenuation of the line photons
in the solar atmosphere. Unless the flare is very close to the solar limb, the 
attenuation of the delayed gamma-ray lines should not be significant (see Hua, 
Ramaty, \& Lingenfelter 1989) as long as the radioactive nuclei do not plunge 
deep in the solar convection zone. The delayed X-ray lines can be more 
significantly attenuated by photoelectric absorption (see below).  

A full knowledge of the delayed 511~keV line flux would require a comprehensive 
calculation of the accelerated particle transport, solar atmospheric depth
distribution of $\beta^+$-emitter production and transport of the emitted
positrons, because (1) the number of 2$\gamma$ line photons produced per 
emitted positron ($f_{511}$) crucially depends on the density, temperature and 
ionization state of the solar annihilation environment \citep{mur05}, (2) 
significant escape of positrons from the annihilation region can occur, and (3)
the line can be attenuated by Compton scattering in the solar atmosphere. 
Here, we simply assumed $f_{511}$=1 (see Kozlovsky et al. 2004; Murphy et al. 
2005) and neglected the line attenuation. 
We see in Tables~4 and 5 that the annihilation line is predicted to be the most 
intense delayed line for hours after the flare end. After $\sim$2~days however, 
its flux can become lower than that of the 846.8~keV line from the decay of 
$^{56}$Co into $^{56}$Fe (see Table~6). We show in Figure~11 the time dependence 
of the 511~keV line flux, for $s$=3.5 and $\Delta t$=10~min, together with the 
contributions of the main radioactive positron emitters to the line production. 
We see that from $\sim$1 to $\sim$14~hours, $^{18}$F is the main source of the 
positrons. Since this radioisotope can be mainly produced by the reaction 
$^{16}$O($^3$He,$p$)$^{18}$F, we suggest that a future detection of the decay 
curve of the solar 511~keV line could provide an independent measurement of the 
flare-accelerated $^3$He abundance. Prompt line measurements have not yet
furnished an unambiguous determination of the fast $^3$He enrichment
(Manzhavidze et al. 1997).

Among the 9 atomic lines listed in Table~1, the most promising appears to be the 
Co K$\alpha$ line at 6.92~keV (Tables~4--5). It is produced from both the decay 
of the isomer $^{58}$Co$^m$ by the conversion of a K-shell electron and 
the decay of $^{57}$Ni by orbital electron capture. Additional important atomic 
lines are the Fe and Ni K$\alpha$ lines at 6.40 and 7.47~keV, respectively. The
X-ray line fluxes shown in Tables~4 and 5 should be taken as upper limits,
however, because photoelectric absorption of the emitted X-rays was not taken 
into account, as it depends on the flare location and model of 
accelerated ion transport. Calculations in the framework of the 
solar magnetic loop model (Hua et al. 1989) showed that the interaction site of
nuclear reactions is expected to be in the lower chromosphere, at solar depths
corresponding to column densities of 10$^{-3}$ to 10$^{-1}$~g cm$^{-2}$. These
results were reinforced by the gamma-ray spectroscopic analyses of 
\citet{ram95}, who showed that the bulk of the nuclear reactions are produced in 
flare regions where the ambient composition is close to coronal, i.e. above the 
photosphere. For such column densities of material of coronal composition, we 
calculated from the photoelectric absorption cross sections of \citet{bal92} 
that the optical depths of 6.92~keV escaping photons are between $\sim$10$^{-3}$ 
and 10$^{-1}$. Thus, the attenuation of this X-ray line is expected to be 
$\lsim$10\% for flares occuring at low heliocentric angles. However, the line 
attenuation can be much higher for flares near the solar limb.

A serious complication to the X-ray line measurements could arise from the
confusion of the radioactivity lines with the intense thermal emission from 
the flare plasma. In particular, this could prevent a detection of the delayed
X-ray lines for hours after the impulsive flaring phase, until the thermal
emission has become sufficiently low. The necessary distinction of thermal and 
nonthermal photons would certainly benefit from an X-ray instrument with high 
spectral resolution, because K$\alpha$ lines from neutral to low-ionized Fe, Co 
or Ni are not expected from thermal plasmas at ionization equilibrium. 
The neutral Co line at 6.92~keV could still be confused, however, with the 
thermal K$\alpha$ line of Fe XXVI at 6.97~keV. The neutral Fe K$\alpha$ line at 
6.40~keV is commonly observed during large solar flares (e.g. Culhane et al. 
1981), as a result of photoionization by flare X-rays and collisional ionization 
by accelerated electrons (e.g. Zarro, Dennis, \& Slater 1992). However, this 
nonthermal line emission is not expected to extend beyond the impulsive phase. 

A near future detection of delayed nuclear gamma-ray lines is perhaps 
more probable. We see in Table~4 that at $t$=30~min after the flare, the
brightest gamma-ray line (after the 511~keV line) is at 1434~keV from the 
$\beta^+$ decay of the isomeric state $^{52}$Mn$^m$ into $^{52}$Cr. The flux of 
this line is predicted to significantly increase as the accelerated particle 
spectrum becomes harder, because it is mainly produced from Fe spallation 
reactions at relatively high energies, $>$10~MeV nucleon$^{-1}$ (Figure~4). For 
$t$$\gsim$3~hours, the radioactivity of $^{52}$Mn$^m$ ($T_{1/2}$=21.1~min) has 
become negligible and the 1434~keV line essentially results from the decay of the 
ground state $^{52}$Mn$^g$ ($T_{1/2}$=5.59~days). Thus, this line remains 
significant for several days after the flare. However, for $t$$\gsim$2~days, the 
most intense line could be at 846.8~keV from $^{56}$Co decay, depending on the 
spectral index $s$ (see Table~6). Additional important gamma-ray lines during 
the first hour are at 1332 and 1792~keV from the radioactivity of $^{60}$Cu. During 
the first two days, one should also look for the line at 931.1~keV from the 
radioactivity of $^{55}$Co and for those at 1369 and 2754~keV from $^{24}$Na decay. 

We now discuss the influence of the accelerated ion composition on the
delayed line emission. In Figure~12, we show calculated fluences of the 846.8 and
1434~keV lines as a function of the accelerated $\alpha/p$ abundance ratio. They
were obtained from the equation
\begin{equation}
\mathcal{F}_l = \int_0^\infty F_l(t) dt ={\mathcal{F}_{4.4+6.1} Q_r f_d I_l^r 
\over Q_{4.4+6.1}}~, 
\end{equation}
where the yields $Q_r$ and $Q_{4.4+6.1}$ were calculated for various proton 
abundances relative to $\alpha$-particles and the other accelerated ions. Thus, 
the predicted fluence variations with accelerated $\alpha/p$ actually 
show the relative contributions of reactions induced by fast protons. The 
fluences decrease for decreasing $\alpha/p$ ratio (i.e. increasing proton 
abundance), because, for $\alpha/p$$\gsim$0.05, the radioisotopes are 
predominantly produced by spallation of accelerated heavy nuclei, whose
abundances are significantly enhanced in impulsive flares, whereas the ambient 
$^{12}$C and $^{16}$O lines largely result from fast proton interactions. This 
effect is less pronounced for $s$=5, because for this very soft spectrum, the 
contribution of $\alpha$-particle reactions to the prompt line emission is more 
important. Obviously, the detection of any delayed line from a solar flare should 
furnish valuable information on the accelerated particle composition 
and energy spectrum. Determination of the accelerated particle composition from
spectroscopy of prompt line emission is difficult.

\section{Discussion}

We have made a detailed evaluation of the nuclear data relevant to the production 
of radioactive line emission in the aftermath of large solar flares. We have
presented updated cross sections for the synthesis of the major radioisotopes by
proton, $^3$He and $\alpha$ reactions, and have provided theoretical
thick-target yields, which allow flux estimates for all the 
major delayed lines at any time after a gamma-ray flare. 

Together with the 846.8~keV line from $^{56}$Co decay, whose importance was
already pointed out by \cite{ram00}, our study has revealed other gamma-ray
lines that appear to be promising for detection, e.g. at 1434 keV from
$^{52}$Mn$^{g,m}$, 1332 and 1792~keV from $^{60}$Cu, 2127 keV from
$^{34}$Cl$^m$, 1369 and 2754~keV from $^{24}$Na, and 931.1~keV from $^{55}$Co. 
The strongest delayed X-ray line is found to be the Co K$\alpha$ 
at 6.92~keV, which is produced from both the decay of the isomer 
$^{58}$Co$^m$ by the conversion of a K-shell electron and the decay of 
$^{57}$Ni by orbital electron capture. Distinguishing this atomic line from 
the thermal X-ray emission can be challenging until the flare plasma has
significantly cooled down. However, a few hours after the flare the thermal 
emission will be gone or significantly reduced and the delayed Co K$\alpha$ 
line will be more easily detected. 

Delayed gamma-ray lines could be detected sooner after the end of the impulsive 
phase, as the prompt nonthermal gamma-ray emission vanishes more rapidly. The 
lines will be very narrow, because the radioactive nuclei are stopped by energy 
losses in the solar atmosphere before they decay. Although generally weaker than 
the main prompt lines, some delayed lines emitted after large flares can have 
fluences within the detection capabilities of the {\it RHESSI} spectrometer or 
future space instruments. Multiple flares originating from the same active region 
of the sun can build up the radioactivity, thus increasing the chance for 
detection. 

However, a major complication to the measurements can arise from the fact that 
the same radioactivity lines can be produced in the instrument and spacecraft 
materials from fast particle interactions. A line of solar origin could 
sometimes be disentangled from the instrumental line at the same energy by 
their different time evolutions. But the bombardment of the satellite by solar 
energetic particles associated with the gamma-ray flare can make this 
selection more difficult.

A positive detection of delayed radioactivity lines, hopefully with {\it
RHESSI}, would certainly provide unique information on the
flare-accelerated particle composition and energy spectrum. In particular, since
the enrichment of the accelerated heavy elements can be the major source of the
radioisotopes, their detection should furnish a valuable measurement of this
enhancement. Thus, a concomitant detection of the two lines at 846.8 and
1434~keV could allow measurement of not only the abundance of accelerated Fe 
ions, but also of their energy spectrum (see Figure~12). 

A future measurement of the decay curve of the electron-positron annihilation line 
or of other delayed gamma-ray lines would be very useful for studying solar 
atmospheric mixing. The lines should be strongly attenuated by Compton scattering 
when the radioactive nuclei plunge deep in the solar interior. The use of 
several radioisotopes with different lifetimes should place constraints on the 
extents and timescales of mixing processes in the outer convection zone. In
addition, the imaging capabilities of {\it RHESSI} could allow measurement of the 
size and development of the radioactive patch on the solar surface. This would 
provide unique information on both the transport of flare-accelerated particles 
and dynamics of solar active regions. It is noteworthy that solar radioactivity 
can be the only way to study flares that had recently occured over the east limb. 

Radioactive nuclei produced in solar flares can also be detected directly if they 
escape from the sun into interplanetary space. At present, only two long-lived 
radioisotopes of solar-flare origin have been identified, $^{14}$C 
($T_{1/2}$=5.7$\times$10$^3$ years, Jull, Lal, \& Donahue 1995) and $^{10}$Be 
($T_{1/2}$=1.51$\times$10$^6$ years, Nishiizumi \& Caffe 2001), from measurements 
of the solar wind implanted in the outer layers of lunar grains. Based on the 
measured abundances relative to calculated average production rates in flares, 
a large part of these radioactive species must be ejected in the solar wind and 
energetic-particle events rather than being mixed into the bulk of the solar 
convection zone. Detection of solar radioactivities with shorter lifetimes, either 
directly in interplanetary space or from their delayed line emission, are 
expected to provide a new insight into the destiny of the nuclei synthesized in 
solar flares. 

\acknowledgments

We would like to thank Amel Belhout for her assistance in the EMPIRE-II
calculations and Jean-Pierre Thibaud for his constructive comments on the
manuscript. B. Kozlovsky would like to thank V. Tatischeff and J. Kiener for
their hospitality at Orsay and acknowledges the Israeli Science Foundation for
support. 




\clearpage
\twocolumn

\begin{figure}
\plotone{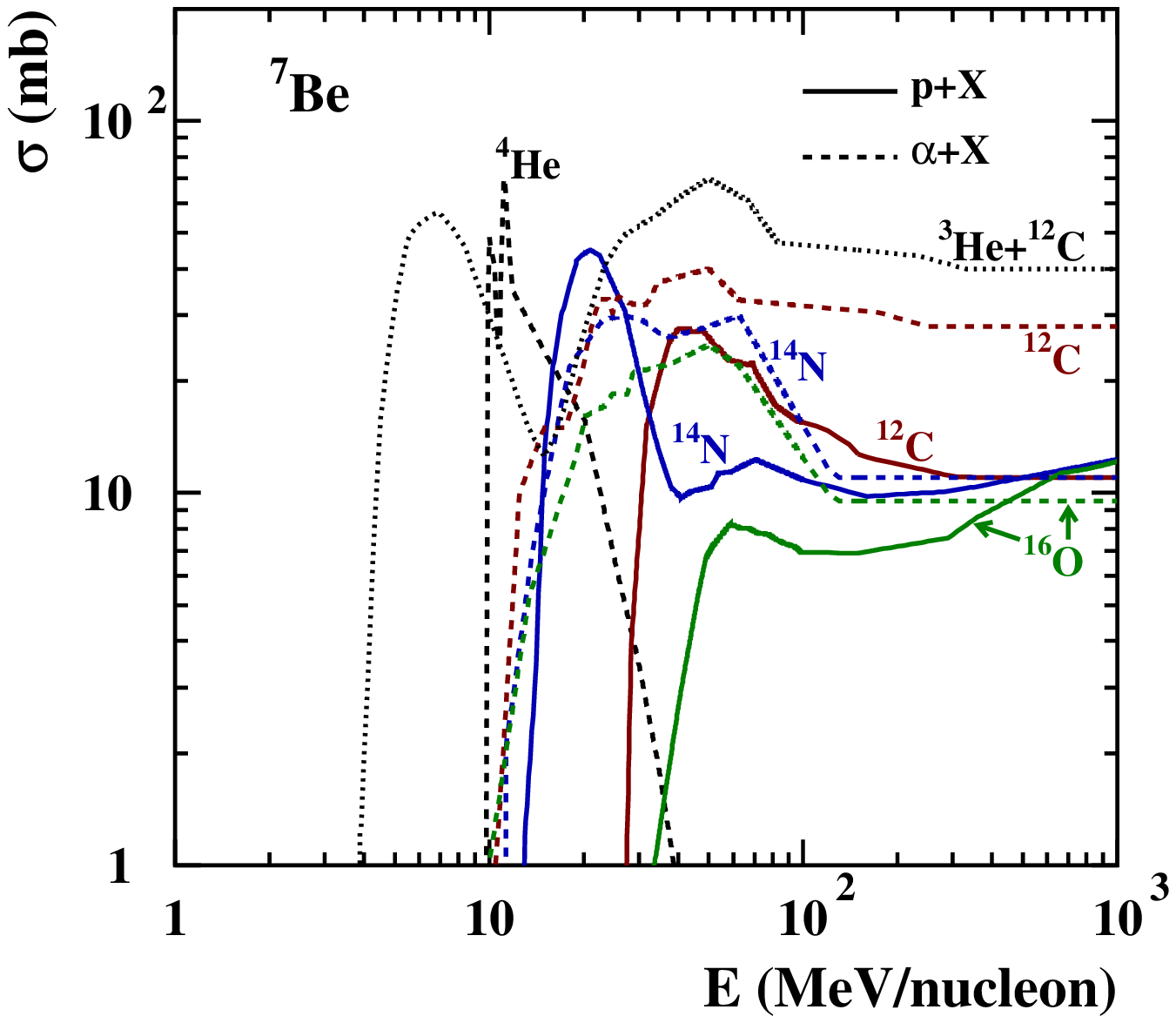}
\figcaption{Cross sections for the production of $^7$Be from $^3$He interactions
with $^{12}$C, the $^4$He($\alpha$,$n$)$^7$Be reaction, and proton and $\alpha$ 
reactions with $^{12}$C, $^{14}$N and $^{16}$O. See the electronic edition of the 
Journal for a color version of this figure.\label{fig1}}
\end{figure}

\begin{figure}
\plotone{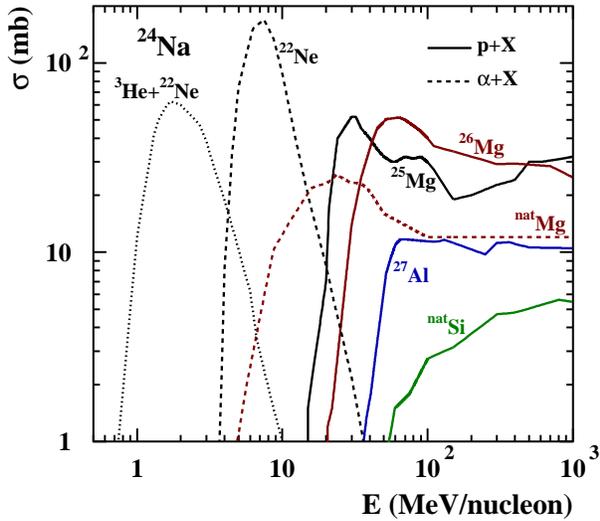}
\figcaption{Cross sections for the production of $^{24}$Na from proton reactions
with $^{25}$Mg, $^{26}$Mg, $^{27}$Al and $^{nat}$Si, $\alpha$ reactions with 
$^{22}$Ne and $^{nat}$Mg, and $^3$He interactions with $^{22}$Ne. See the electronic 
edition of the Journal for a color version of this figure.\label{fig2}}
\end{figure}

\begin{figure}
\plotone{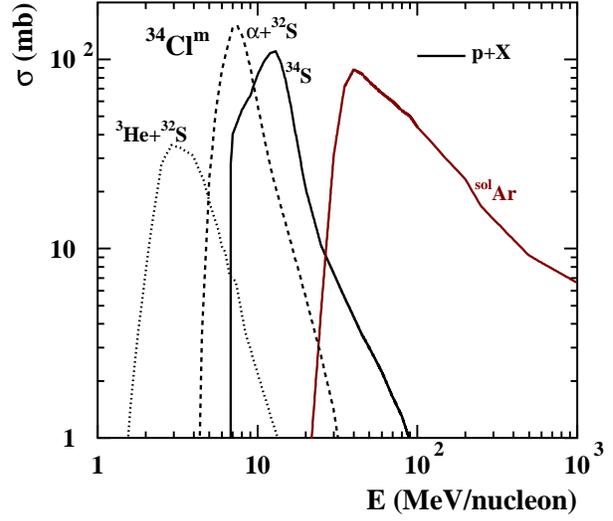}
\figcaption{Cross sections for the production of $^{34}$Cl$^m$ from proton reactions
with $^{34}$S and Ar of solar isotopic composition (see text), and $^3$He and 
$\alpha$ reactions with $^{32}$S. See the electronic edition of the 
Journal for a color version of this figure.\label{fig3}}
\end{figure}

\begin{figure}
\plotone{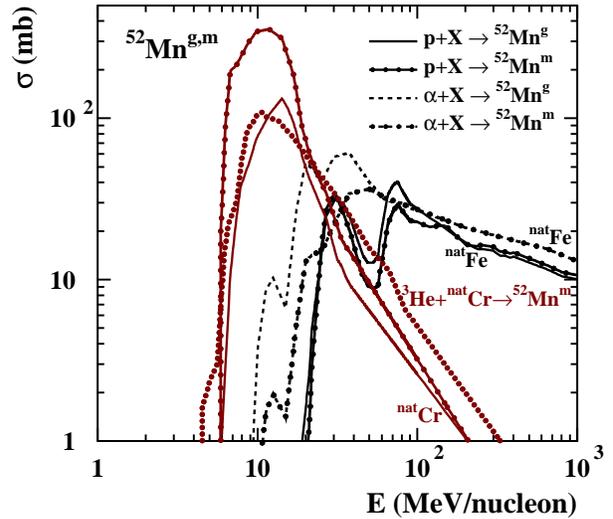}
\figcaption{Cross sections for the production of $^{52}$Mn$^g$ and $^{52}$Mn$^m$
from proton reactions with $^{nat}$Cr and $^{nat}$Fe, $^3$He reactions with 
$^{nat}$Cr and $\alpha$ reactions with $^{nat}$Fe. See the electronic edition of the 
Journal for a color version of this figure.\label{fig4}}
\end{figure}

\begin{figure}
\plotone{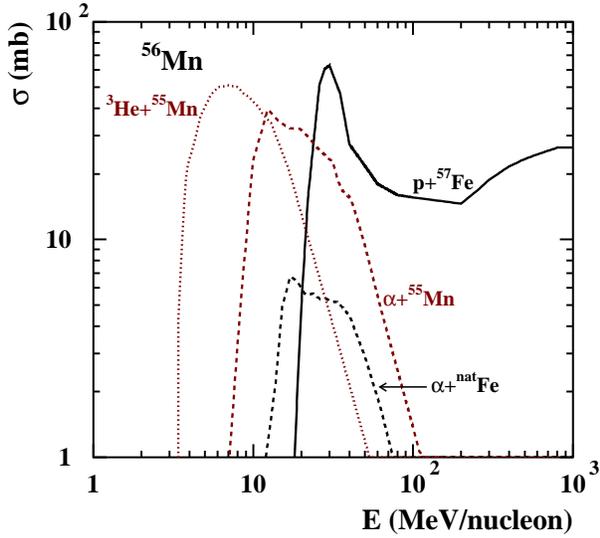}
\figcaption{Cross sections for the production of $^{56}$Mn from $^3$He and $\alpha$ 
reactions with $^{55}$Mn, $\alpha$ reactions with $^{nat}$Fe, and the
$^{57}$Fe($p$,2$p$)$^{56}$Mn reaction. See the electronic edition of the Journal for 
a color version of this figure.\label{fig5}}
\end{figure}

\begin{figure}
\plotone{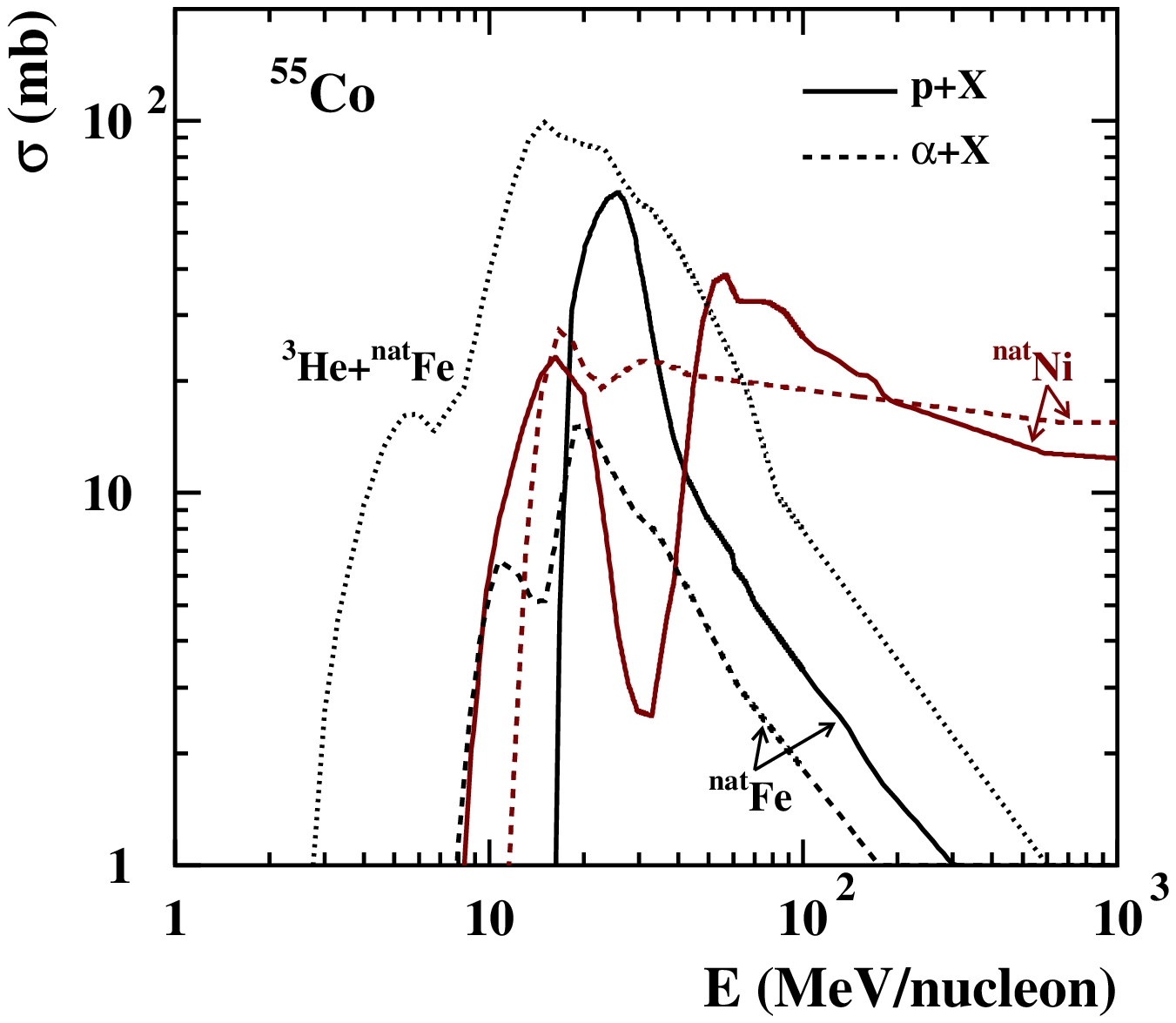}
\figcaption{Cross sections for the production of $^{55}$Co from proton and $\alpha$ 
reactions with $^{nat}$Fe and $^{nat}$Ni, and $^3$He reactions with $^{nat}$Fe. See 
the electronic edition of the Journal for a color version of this 
figure.\label{fig6}}
\end{figure}

\begin{figure}
\plotone{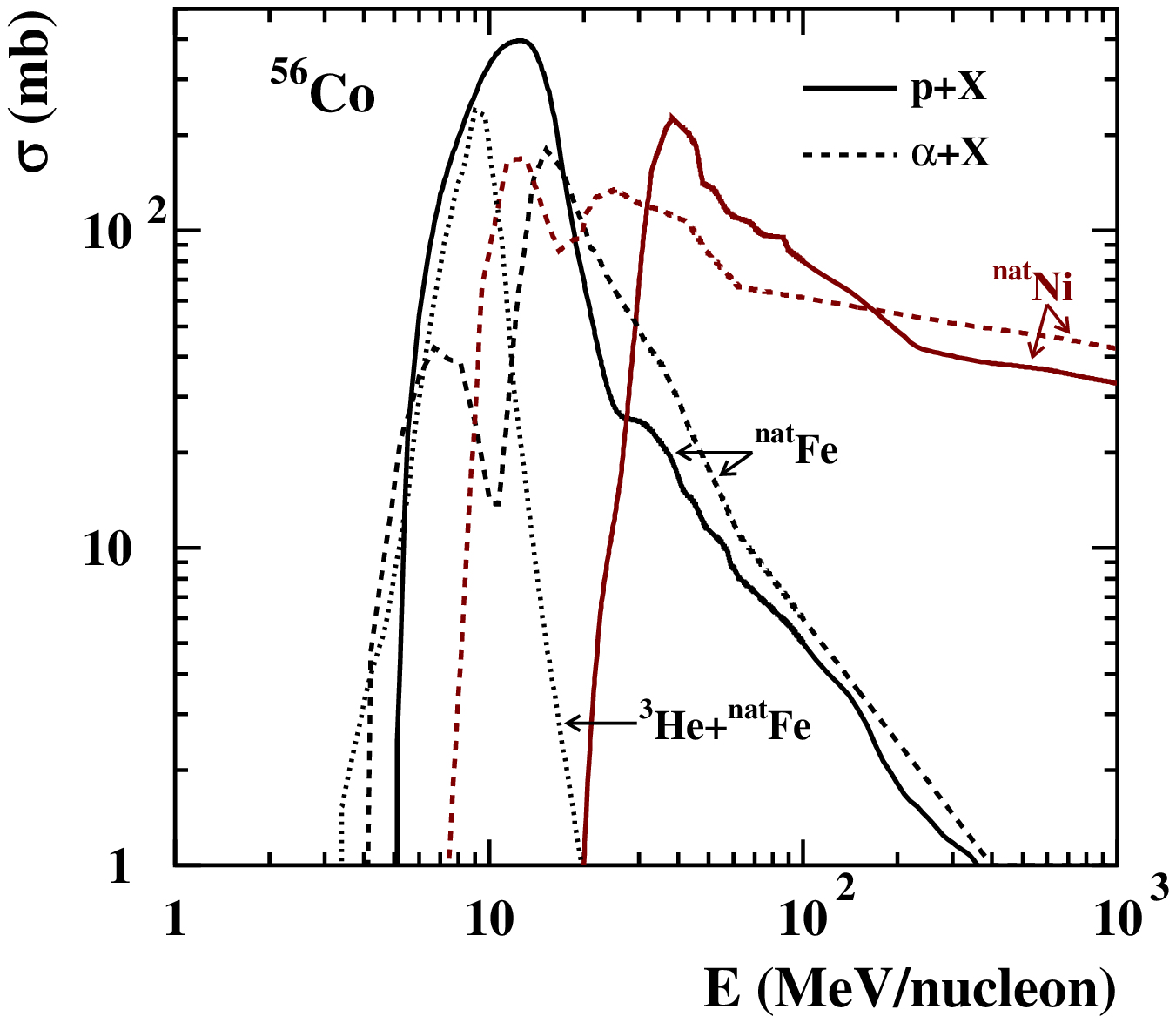}
\figcaption{Cross sections for the production of $^{56}$Co from proton and $\alpha$ 
reactions with $^{nat}$Fe and $^{nat}$Ni, and $^3$He reactions with $^{nat}$Fe. See 
the electronic edition of the Journal for a color version of this
figure.\label{fig7}}
\end{figure}

\begin{figure}
\plotone{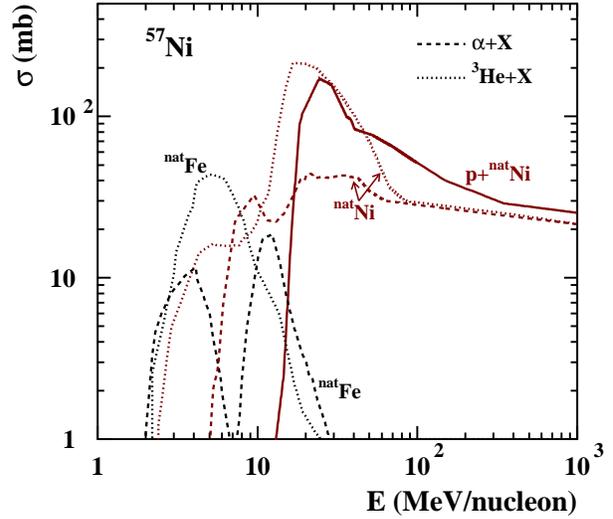}
\figcaption{Cross sections for the production of $^{57}$Ni from $^3$He and $\alpha$ 
reactions with $^{nat}$Fe and $^{nat}$Ni, and proton reactions with $^{nat}$Ni. See 
the electronic edition of the Journal for a color version of this
figure.\label{fig8}}
\end{figure}

\begin{figure}
\begin{center}
\plotone{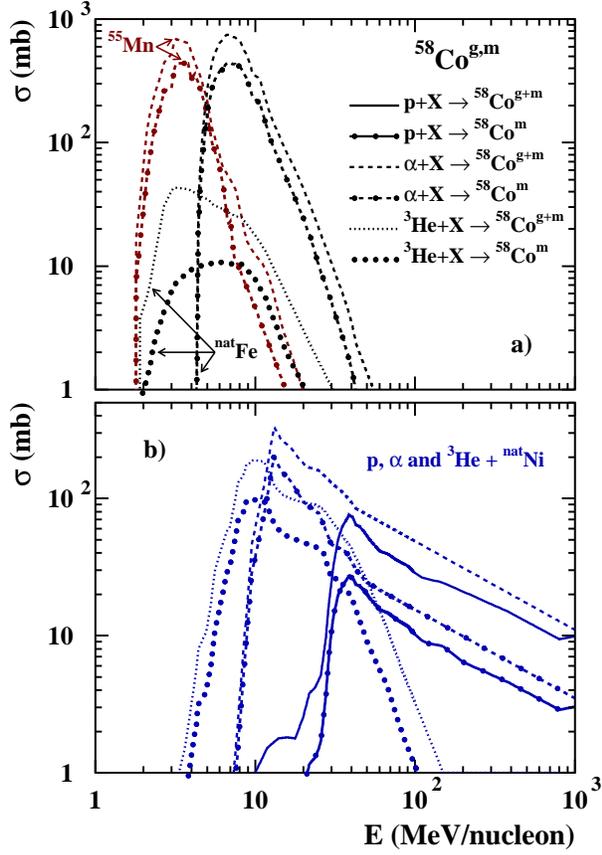}
\end{center}
\figcaption{Cross sections for the total production of $^{58}$Co (i.e. both
$^{58}$Co$^g$ and $^{58}$Co$^m$) and of $^{58}$Co$^m$ from (a) $\alpha$ 
reactions with $^{55}$Mn and $^{nat}$Fe, as well as $^3$He reactions with 
$^{nat}$Fe, and (b) proton, $^3$He and $\alpha$ reactions with $^{nat}$Ni. See 
the electronic edition of the Journal for a color version of this
figure.\label{fig9}}
\end{figure}

\begin{figure}
\plotone{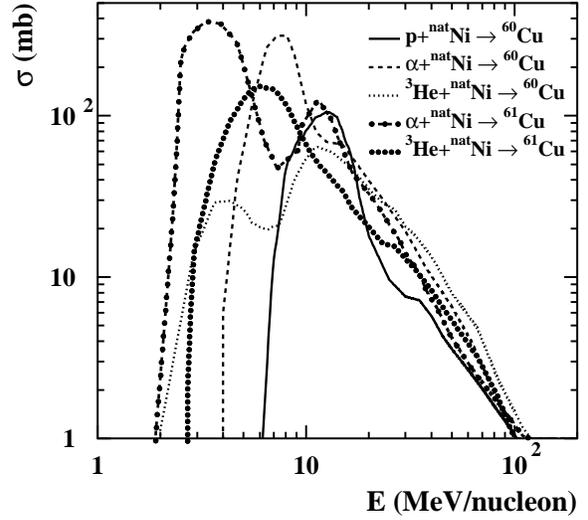}
\figcaption{Cross sections for the production of $^{60}$Cu and $^{61}$Cu from 
proton, $^3$He and $\alpha$ reactions with $^{nat}$Ni.\label{fig10}}
\end{figure}

\begin{figure}
\plotone{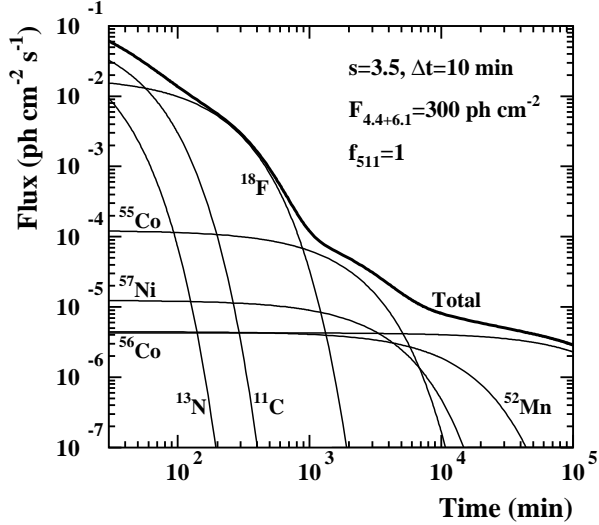}
\figcaption{Time dependence of the 511 keV line flux with the contributions 
of the main $\beta^+$-emitters, for a spectral index $s$=3.5 and a flare 
duration $\Delta t$=10~min. The calculations are normalized to a total 
fluence of 300 photons cm$^{-2}$ emitted during the gamma-ray flare in the 
sum of the 4.44 and 6.13 MeV ambient $^{12}$C and $^{16}$O deexcitation 
lines. The positron to annihilation-line photon conversion factor 
$f_{511}$=1. Line attenuation in the solar atmosphere is not 
taken into account.\label{fig11}}
\end{figure}

\begin{figure}
\plotone{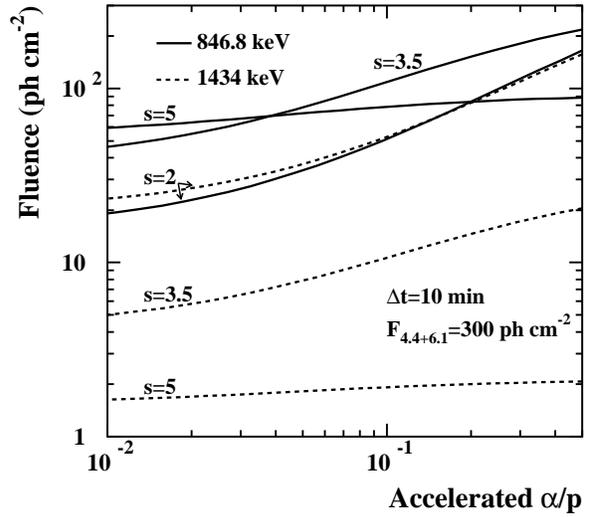}
\figcaption{Total fluences (eq.~[9]) of the delayed lines at 846.8~keV (solid 
curves) and 1434~keV (dashed curves) as a function of the accelerated 
$\alpha/p$ abundance ratio (see text), for $s$=2, 3.5 and 5. The calculations 
are normalized to a total fluence of 300 photons cm$^{-2}$ emitted during the 
gamma-ray flare in the sum of the 4.44 and 6.13 MeV ambient $^{12}$C and 
$^{16}$O deexcitation lines.\label{fig12}}
\end{figure}


\clearpage
\onecolumn

\begin{deluxetable}{lll}
\footnotesize
\tablewidth{0pt}
\tablecaption{Radioactive X- and gamma-ray line emitters ordered by 
lifetime\label{tab1}}
\tablehead{
\colhead{Isotope} & \colhead{Half-life\tablenotemark{a}} & 
\colhead{Photon energy (keV) and intensity (\%)\tablenotemark{b}} }
\startdata
$^{13}$N   & 9.965 (4)  m  & 511 (99.8) \\
$^{11}$C   & 20.385 (20) m  & 511 (99.8) \\
$^{52}$Mn$^m$ & 21.1 (2) m  & 511 (98.2), 1434 (99.8) \\
$^{60}$Cu  & 23.7 (4) m  & 7.47 (2.5), 511 (93.0), 826.4 (21.7), 1332 (88.0), 
1792 (45.4) \\
$^{34}$Cl$^m$ & 32.00 (4)   m  & 146.4 (40.5), 511 (54.3), 1177 (14.1), 2127 (42.8), 
3304 (12.3) \\
$^{47}$V   & 32.6 (3)   m  & 511 (96.5) \\
$^{63}$Zn  & 38.47 (5)  m  & 8.04 (2.5), 511 (92.7) \\
$^{49}$Cr  & 42.3 (1)   m  & 4.95 (2.5), 62.3 (16.4), 90.6 (53.2), 152.9 (30.3), 
511 (92.6) \\
$^{51}$Mn  & 46.2 (1)   m  & 511 (97.1) \\
$^{18}$F   & 109.77 (5) m  & 511 (96.7) \\ 
$^{56}$Mn  & 2.5789 (1) h  & 846.8 (98.9), 1811 (27.2), 2113 (14.3) \\
$^{45}$Ti  & 184.8 (5)  m  & 4.09 (2.3), 511 (84.8) \\
$^{61}$Cu  & 3.333 (5)  h  & 7.47 (12.6), 283.0 (12.2), 511 (61.0), 656.0 (10.8) \\
$^{43}$Sc  & 3.891 (12)  h  & 372.9 (22.5), 511 (88.1) \\
$^{44}$Sc  & 3.97  (4)  h  & 511 (94.3), 1157 (99.9) \\
$^{52}$Fe  & 8.275 (8) h  & 5.90 (11.2), 168.7 (99.2), 511 (55.5) \\
$^{58}$Co$^m$ & 9.04 (11)   h  & 6.92 (23.8) \\
$^{24}$Na  & 14.9590 (12)  h  & 1369 (100), 2754 (99.9) \\
$^{55}$Co  & 17.53 (3)  h  & 6.40 (6.5), 477.2 (20.2), 511 (76.0), 931.1 (75.0), 
1408 (16.9) \\
$^{57}$Ni  & 35.60 (6)  h  & 6.92 (16.7), 127.2 (16.7), 511 (43.6), 1378 (81.7), 
1920 (12.3) \\
$^{52}$Mn$^g$  & 5.591 (3)  d  & 5.41 (15.5), 511 (29.6), 744.2 (90.0), 935.5 (94.5), 
1434 (100) \\
$^{48}$V   & 15.9735 (25) d  & 4.51 (8.6), 511 (50.3), 983.5 (100), 1312 (97.5) \\
$^{7}$Be   & 53.22 (6)  d  & 477.6 (10.4) \\
$^{58}$Co$^g$  & 70.86  (6) d  & 6.40 (23.0), 511 (14.9), 810.8 (99.4) \\
$^{56}$Co  & 77.233 (27) d  & 6.40 (21.8), 511 (19.0), 846.8 (99.9), 1038 (14.2), \\
 & & 1238 (66.9), 1771 (15.5), 2598 (17.3) \\
\enddata
\tablenotetext{a}{The half-life uncertainty is indicated by the number in
parenthesis, which represents the uncertainty in the least significant
digit(s). The units are: m=minute, h=hour and d=day. The decay data are 
extracted from the NuDat database, URL http://www.nndc.bnl.gov/nudat2/}
\tablenotetext{b}{ Number of line photons emitted per 100 radioactive decays,
except for the 511 keV positron annihilation line for which the given intensity 
is for the positron production. All K$\alpha$-X- and gamma-ray lines with 
intensities greater than 2 and 10\%, respectively, were considered.}
\end{deluxetable}

\begin{deluxetable}{ll}
\footnotesize
\tablewidth{0pt}
\tablecaption{Radioisotope production reactions\label{tab2}}
\tablehead{
\colhead{Isotope} & \colhead{Reactions} }
\startdata
$^{7}$Be & $^4$He($\alpha$,$n$), $^{12}$C($p$,$x$), $^{12}$C($^3$He,$x$), 
$^{12}$C($\alpha$,$x$), $^{14}$N($p$,$x$), $^{14}$N($\alpha$,$x$),
$^{16}$O($p$,$x$), $^{16}$O($\alpha$,$x$) \\
$^{24}$Na & $^{22}$Ne($^3$He,$p$), $^{22}$Ne($\alpha$,$pn$),
$^{25}$Mg($p$,2$p$), $^{26}$Mg($p$,2$pn$), $^{nat}$Mg($\alpha$,$x$), 
$^{27}$Al($p$,3$pn$)\tablenotemark{a}, $^{nat}$Si($p$,$x$) \\
$^{34}$Cl$^m$ & $^{32}$S($^3$He,$p$), $^{32}$S($\alpha$,$pn$),
$^{34}$S($p$,$n$), $^{sol}$Ar($p$,$x$)\tablenotemark{b}\\
$^{52}$Mn$^{g,m}$ & $^{nat}$Cr($p$,$x$),
$^{nat}$Cr($^3$He,$x$)$^{52}$Mn$^{m~}$\tablenotemark{c},
$^{nat}$Fe($p$,$x$), $^{nat}$Fe($\alpha$,$x$) \\
$^{56}$Mn & $^{55}$Mn($^3$He,2$p$), $^{55}$Mn($\alpha$,2$pn$), 
$^{57}$Fe($p$,2$p$), $^{nat}$Fe($\alpha$,$x$) \\
$^{55}$Co & $^{nat}$Fe($p$,$x$), $^{nat}$Fe($^3$He,$x$), 
$^{nat}$Fe($\alpha$,$x$), $^{nat}$Ni($p$,$x$), $^{nat}$Ni($\alpha$,$x$) \\
$^{56}$Co & $^{nat}$Fe($p$,$x$), $^{nat}$Fe($^3$He,$x$), 
$^{nat}$Fe($\alpha$,$x$), $^{nat}$Ni($p$,$x$), $^{nat}$Ni($\alpha$,$x$) \\
$^{57}$Ni & $^{nat}$Fe($p$,$x$), $^{nat}$Fe($^3$He,$x$), 
$^{nat}$Fe($\alpha$,$x$), $^{nat}$Ni($p$,$x$), $^{nat}$Ni($^3$He,$x$), 
$^{nat}$Ni($\alpha$,$x$) \\
$^{58}$Co$^{g,m}$ & $^{55}$Mn($\alpha$,$n$), $^{nat}$Fe($^3$He,$x$), 
$^{nat}$Fe($\alpha$,$x$), $^{nat}$Ni($p$,$x$), $^{nat}$Ni($^3$He,$x$), 
$^{nat}$Ni($\alpha$,$x$) \\
$^{60}$Cu & $^{nat}$Ni($p$,$x$), $^{nat}$Ni($^3$He,$x$), 
$^{nat}$Ni($\alpha$,$x$) \\
$^{61}$Cu & $^{nat}$Ni($^3$He,$x$), $^{nat}$Ni($\alpha$,$x$) \\
\enddata
\tablenotetext{a}{Here and in the following, the notation "$xpn$" in the reaction
output channel denotes $x$$p$+$n$, ($x$-1)$p$+$d$ or ($x$-2)$p$+$^3$He
(if $x$$\geq$2).}
\tablenotetext{b}{The notation $^{sol}$Ar means Ar of solar isotopic composition
(see text).}
\tablenotetext{c}{The contribution of $^3$He+$^{nat}$Cr collisions to $^{52}$Mn$^g$
production was safely neglected.}
\end{deluxetable}

\begin{deluxetable}{lccccc}
\footnotesize
\tablewidth{0pt}
\tablecaption{Radioisotope production yields\label{tab3}}
\tablehead{ \colhead{Isotope} & 
\multicolumn{3}{c}{Production yields\tablenotemark{a}} & 
\colhead{$f_d$\tablenotemark{b}} & 
\colhead{Ref.\tablenotemark{c}} \\  \cline{2-4}
& $s$=3.5 & $s$=2 & $s$=5}
\startdata
$^{13}$N     & 1.43$\times$10$^{-6}$ & 1.49$\times$10$^{-5}$ & 8.89$\times$10$^{-7}$  & 0.72 & 1 \\
$^{11}$C     & 2.98$\times$10$^{-6}$ & 5.80$\times$10$^{-5}$ & 5.12$\times$10$^{-6}$  & 0.85 & 1 \\
$^{52}$Mn$^m$& 1.88$\times$10$^{-7}$ & 7.42$\times$10$^{-6}$ & 2.15$\times$10$^{-8}$  & 0.85 & 2 \\
$^{60}$Cu    & 4.24$\times$10$^{-8}$ & 1.49$\times$10$^{-7}$ & 1.70$\times$10$^{-8}$  & 0.87 & 2 \\
$^{34}$Cl$^m$& 4.09$\times$10$^{-8}$ & 4.69$\times$10$^{-7}$ & 1.96$\times$10$^{-8}$  & 0.90 & 2 \\
$^{47}$V     & 2.34$\times$10$^{-8}$ & 1.58$\times$10$^{-6}$ & 5.95$\times$10$^{-10}$ & 0.90 & 1 \\
$^{63}$Zn    & 1.02$\times$10$^{-8}$ & 7.63$\times$10$^{-9}$ & 1.58$\times$10$^{-8}$  & 0.92 & 1 \\
$^{49}$Cr    & 1.71$\times$10$^{-8}$ & 1.84$\times$10$^{-6}$ & 2.27$\times$10$^{-10}$ & 0.92 & 1 \\
$^{51}$Mn    & 4.08$\times$10$^{-8}$ & 1.94$\times$10$^{-6}$ & 2.39$\times$10$^{-9}$  & 0.93 & 1 \\
$^{18}$F     & 2.99$\times$10$^{-6}$ & 2.66$\times$10$^{-5}$ & 4.99$\times$10$^{-6}$  & 0.97 & 1 \\
$^{56}$Mn    & 1.01$\times$10$^{-8}$ & 2.52$\times$10$^{-7}$ & 1.02$\times$10$^{-9}$  & 0.98 & 2 \\
$^{45}$Ti    & 1.46$\times$10$^{-0}$ & 1.68$\times$10$^{-8}$ & 1.69$\times$10$^{-12}$ & 0.98 & 1 \\
$^{61}$Cu    & 4.55$\times$10$^{-8}$ & 4.50$\times$10$^{-8}$ & 7.55$\times$10$^{-8}$  & 0.98 & 2 \\
$^{43}$Sc    & 5.67$\times$10$^{-9}$ & 1.94$\times$10$^{-8}$ & 6.22$\times$10$^{-9}$  & 0.99 & 1 \\
$^{44}$Sc    & 4.29$\times$10$^{-0}$ & 2.68$\times$10$^{-8}$ & 2.18$\times$10$^{-11}$ & 0.99 & 1 \\
$^{52}$Fe    & 1.38$\times$10$^{-9}$ & 9.31$\times$10$^{-8}$ & 3.35$\times$10$^{-11}$ & 0.99 & 1 \\
$^{58}$Co$^m$& 6.15$\times$10$^{-7}$ & 1.30$\times$10$^{-6}$ & 3.92$\times$10$^{-7}$  & 0.99 & 2 \\
$^{24}$Na    & 9.94$\times$10$^{-8}$ & 3.59$\times$10$^{-6}$ & 3.64$\times$10$^{-8}$  & 1 & 2 \\
$^{55}$Co    & 2.34$\times$10$^{-7}$ & 3.18$\times$10$^{-6}$ & 2.59$\times$10$^{-8}$  & 1 & 2 \\
$^{57}$Ni    & 8.41$\times$10$^{-8}$ & 8.76$\times$10$^{-7}$ & 7.78$\times$10$^{-8}$  & 1 & 2 \\
$^{52}$Mn$^g$& 1.67$\times$10$^{-7}$ & 7.24$\times$10$^{-6}$ & 1.01$\times$10$^{-8}$  & 1 & 2 \\
$^{48}$V     & 4.54$\times$10$^{-8}$ & 2.89$\times$10$^{-6}$ & 2.22$\times$10$^{-9}$  & 1 & 1 \\
$^{7}$Be     & 9.07$\times$10$^{-6}$ & 6.33$\times$10$^{-5}$ & 2.07$\times$10$^{-6}$  & 1 & 2 \\
$^{58}$Co$^g$& 1.01$\times$10$^{-6}$ & 2.33$\times$10$^{-6}$ & 6.54$\times$10$^{-7}$  & 1 & 2 \\
$^{56}$Co    & 3.46$\times$10$^{-6}$ & 1.40$\times$10$^{-5}$ & 1.21$\times$10$^{-6}$  & 1 & 2 \\
4.44 MeV     & 2.73$\times$10$^{-6}$ & 1.38$\times$10$^{-5}$ & 2.34$\times$10$^{-6}$  & - & 2 \\
6.13 MeV     & 2.04$\times$10$^{-6}$ & 1.20$\times$10$^{-5}$ & 1.76$\times$10$^{-6}$  & - & 2 \\
\enddata
\tablenotetext{a}{Nondimensional (eq.~[3]). The calculations are normalized to 
unit number of accelerated protons of energies greater than 5~MeV impinging on 
the solar thick target. The last two lines of the table give the production 
yields of the 4.44 and 6.13 MeV deexcitation lines from solar ambient $^{12}$C 
and $^{16}$O, respectively.}
\tablenotetext{b}{Factor which takes into account the decay of the
radioactive isotopes during the flare, assuming a flare duration of 10~min and 
a steady-state radioisotope production (see eq.~[7]).}
\tablenotetext{c}{Source of the radioisotope production cross sections-- (1)
\citet{koz87,koz04}; (2) this work.}
\end{deluxetable}

\begin{deluxetable}{llccc}
\footnotesize
\tablewidth{0pt}
\tablecaption{Delayed X- and gamma-ray line fluxes at $t$=30~min\label{tab4}}
\tablehead{ \colhead{Energy} & 
\colhead{Parent} & 
\multicolumn{3}{c}{Line flux (photons cm$^{-2}$ s$^{-1}$)\tablenotemark{b}} \\  
\cline{3-5} \colhead{(keV)} & \colhead{nucleus\tablenotemark{a}} & 
$s$=3.5 & $s$=2 & $s$=5}
\startdata
511&$^{11}$C, $^{18}$F, $^{13}$N... & 6.09$\times$10$^{-2}$ & 1.85$\times$10$^{-1}$ & 1.02$\times$10$^{-1}$ \\
 1434& $^{52}$Mn$^m$, $^{52}$Mn$^g$ & 2.07$\times$10$^{-3}$ & 1.51$\times$10$^{-2}$ & 2.74$\times$10$^{-4}$ \\
 1332    &     $^{60}$Cu            & 4.12$\times$10$^{-4}$ & 2.67$\times$10$^{-4}$ & 1.93$\times$10$^{-4}$ \\
 1792    &     $^{60}$Cu            & 2.13$\times$10$^{-4}$ & 1.38$\times$10$^{-4}$ & 9.95$\times$10$^{-5}$ \\
    6.92 & $^{58}$Co$^m$, $^{57}$Ni & 1.92$\times$10$^{-4}$ & 8.25$\times$10$^{-5}$ & 1.44$\times$10$^{-4}$ \\
 2127    &     $^{34}$Cl$^m$        & 1.87$\times$10$^{-4}$ & 3.95$\times$10$^{-4}$ & 1.04$\times$10$^{-4}$ \\
  146.4  &     $^{34}$Cl$^m$        & 1.76$\times$10$^{-4}$ & 3.74$\times$10$^{-4}$ & 9.86$\times$10$^{-5}$ \\
  931.1  &     $^{55}$Co            & 1.18$\times$10$^{-4}$ & 2.98$\times$10$^{-4}$ & 1.53$\times$10$^{-5}$ \\
  826.4  &     $^{60}$Cu            & 1.02$\times$10$^{-4}$ & 6.59$\times$10$^{-5}$ & 4.75$\times$10$^{-5}$ \\
   90.6  &     $^{49}$Cr            & 8.80$\times$10$^{-5}$ & 1.75$\times$10$^{-3}$ & 1.36$\times$10$^{-6}$ \\
 1369    &     $^{24}$Na            & 7.84$\times$10$^{-5}$ & 5.23$\times$10$^{-4}$ & 3.33$\times$10$^{-5}$ \\
 2754    &     $^{24}$Na            & 7.83$\times$10$^{-5}$ & 5.22$\times$10$^{-4}$ & 3.33$\times$10$^{-5}$ \\
  846.8  & $^{56}$Mn, $^{56}$Co     & 6.29$\times$10$^{-5}$ & 2.02$\times$10$^{-4}$ & 1.39$\times$10$^{-5}$ \\
 1177    &     $^{34}$Cl$^m$        & 6.14$\times$10$^{-5}$ & 1.30$\times$10$^{-4}$ & 3.43$\times$10$^{-5}$ \\
 3304    &     $^{34}$Cl$^m$        & 5.36$\times$10$^{-5}$ & 1.14$\times$10$^{-4}$ & 3.00$\times$10$^{-5}$ \\
  152.9  &     $^{49}$Cr            & 5.01$\times$10$^{-5}$ & 9.98$\times$10$^{-4}$ & 7.76$\times$10$^{-7}$ \\
  477.2  &     $^{55}$Co            & 3.19$\times$10$^{-5}$ & 8.02$\times$10$^{-5}$ & 4.11$\times$10$^{-6}$ \\
    7.47 & $^{61}$Cu, $^{60}$Cu     & 3.02$\times$10$^{-5}$ & 1.10$\times$10$^{-5}$ & 4.11$\times$10$^{-5}$ \\
   62.3  &     $^{49}$Cr            & 2.71$\times$10$^{-5}$ & 5.40$\times$10$^{-4}$ & 4.20$\times$10$^{-7}$ \\
 1408    &     $^{55}$Co            & 2.67$\times$10$^{-5}$ & 6.71$\times$10$^{-5}$ & 3.44$\times$10$^{-6}$ \\
 1378    &     $^{57}$Ni            & 2.31$\times$10$^{-5}$ & 4.45$\times$10$^{-5}$ & 2.49$\times$10$^{-5}$ \\
  283.0  &     $^{61}$Cu            & 1.79$\times$10$^{-5}$ & 3.26$\times$10$^{-6}$ & 3.45$\times$10$^{-5}$ \\
6.40&$^{55}$Co, $^{56}$Co, $^{58}$Co$^g$ & 1.69$\times$10$^{-5}$ & 3.02$\times$10$^{-5}$ & 4.57$\times$10$^{-6}$ \\
  656.0  &     $^{61}$Cu            & 1.58$\times$10$^{-5}$ & 2.89$\times$10$^{-6}$ & 3.05$\times$10$^{-5}$ \\
 1238    &     $^{56}$Co            & 1.51$\times$10$^{-5}$ & 1.13$\times$10$^{-5}$ & 6.16$\times$10$^{-6}$ \\
  935.5  &     $^{52}$Mn$^g$        & 1.42$\times$10$^{-5}$ & 1.14$\times$10$^{-4}$ & 1.00$\times$10$^{-6}$ \\
  744.2  &     $^{52}$Mn$^g$        & 1.35$\times$10$^{-5}$ & 1.08$\times$10$^{-4}$ & 9.54$\times$10$^{-7}$ \\
 1811    &     $^{56}$Mn            & 1.11$\times$10$^{-5}$ & 5.08$\times$10$^{-5}$ & 1.30$\times$10$^{-6}$ \\
\enddata                                 
\tablenotetext{a}{For lines having multiple progenitors, the latter are 
given in decreasing order of their contribution to the line flux. For the 511~keV 
positron annihilation line, only the three main $\beta^+$-emitters are 
indicated.}
\tablenotetext{b}{The calculations are normalized to a total fluence of 300
photons cm$^{-2}$ emitted during the gamma-ray flare in the sum of the 4.44 
and 6.13 MeV ambient $^{12}$C and $^{16}$O deexcitation lines. Decay of the 
radioisotope during the flare is taken into account with the multiplying factor 
$f_d$ given in Table~2. The 511~keV line fluxes are obtained with a positron to 
annihilation-line photon conversion factor $f_{511}$=1. The fluxes of the 6.92, 
7.47 and 6.40~keV lines should be taken as upper limits, because photoelectric 
absorption was not taken into account (see text). Only the lines whose flux is 
$>$10$^{-5}$ photons cm$^{-2}$ s$^{-1}$ for $s$=3.5 are shown.}
\end{deluxetable}

\begin{deluxetable}{llccc}
\footnotesize
\tablewidth{0pt}
\tablecaption{Delayed X- and gamma-ray line fluxes at $t$=3 hours\label{tab5}}
\tablehead{ \colhead{Energy} & 
\colhead{Parent} & 
\multicolumn{3}{c}{Line flux (photons cm$^{-2}$ s$^{-1}$)\tablenotemark{a}} \\  
\cline{3-5} \colhead{(keV)} & \colhead{nucleus\tablenotemark{a}} & 
$s$=3.5 & $s$=2 & $s$=5}
\startdata
 511 & $^{18}$F, $^{11}$C, $^{55}$Co... & 6.44$\times$10$^{-3}$ & 1.17$\times$10$^{-2}$ & 1.21$\times$10$^{-2}$ \\
    6.92 & $^{58}$Co$^m$, $^{57}$Ni     & 1.59$\times$10$^{-4}$ & 6.92$\times$10$^{-5}$ & 1.19$\times$10$^{-4}$ \\
  931.1  &     $^{55}$Co                & 1.07$\times$10$^{-4}$ & 2.70$\times$10$^{-4}$ & 1.38$\times$10$^{-5}$ \\
 1369    &     $^{24}$Na                & 6.98$\times$10$^{-5}$ & 4.66$\times$10$^{-4}$ & 2.97$\times$10$^{-5}$ \\
 2754    &     $^{24}$Na                & 6.97$\times$10$^{-5}$ & 4.65$\times$10$^{-4}$ & 2.97$\times$10$^{-5}$ \\
  846.8  & $^{56}$Co, $^{56}$Mn         & 4.31$\times$10$^{-5}$ & 1.11$\times$10$^{-4}$ & 1.16$\times$10$^{-5}$ \\
 1434    & $^{52}$Mn$^m$, $^{52}$Mn$^g$ & 2.97$\times$10$^{-5}$ & 2.27$\times$10$^{-4}$ & 3.03$\times$10$^{-6}$ \\
  477.2  &     $^{55}$Co                & 2.89$\times$10$^{-5}$ & 7.26$\times$10$^{-5}$ & 3.72$\times$10$^{-6}$ \\
 1408    &     $^{55}$Co                & 2.42$\times$10$^{-5}$ & 6.08$\times$10$^{-5}$ & 3.11$\times$10$^{-6}$ \\
 1378    &     $^{57}$Ni                & 2.20$\times$10$^{-5}$ & 4.24$\times$10$^{-5}$ & 2.37$\times$10$^{-5}$ \\
6.40&$^{55}$Co, $^{56}$Co, $^{58}$Co$^g$& 1.59$\times$10$^{-5}$ & 2.78$\times$10$^{-5}$ & 4.45$\times$10$^{-6}$ \\
 1238    &     $^{56}$Co                & 1.51$\times$10$^{-5}$ & 1.13$\times$10$^{-5}$ & 6.15$\times$10$^{-6}$ \\
  935.5  &     $^{52}$Mn$^g$            & 1.40$\times$10$^{-5}$ & 1.12$\times$10$^{-4}$ & 9.89$\times$10$^{-7}$ \\
  744.2  &     $^{52}$Mn$^g$            & 1.33$\times$10$^{-5}$ & 1.07$\times$10$^{-4}$ & 9.42$\times$10$^{-7}$ \\
    7.47 & $^{61}$Cu, $^{60}$Cu         & 1.11$\times$10$^{-5}$ & 2.10$\times$10$^{-6}$ & 2.12$\times$10$^{-5}$ \\
  283    &     $^{61}$Cu                & 1.06$\times$10$^{-5}$ & 1.94$\times$10$^{-6}$ & 2.05$\times$10$^{-5}$ \\
\enddata
\tablenotetext{a}{Same as Table~4.}
\end{deluxetable}

\begin{deluxetable}{llccc}
\footnotesize
\tablewidth{0pt}
\tablecaption{Delayed X- and gamma-ray line fluxes at $t$=3 days\label{tab6}}
\tablehead{ \colhead{Energy} & 
\colhead{Parent} & 
\multicolumn{3}{c}{Line flux (photons cm$^{-2}$ s$^{-1}$)\tablenotemark{a}} \\  
\cline{3-5} \colhead{(keV)} & \colhead{nucleus\tablenotemark{a}} & 
$s$=3.5 & $s$=2 & $s$=5}
\startdata
  846.8 & $^{56}$Co, $^{56}$Mn        & 2.20$\times$10$^{-5}$ & 1.64$\times$10$^{-5}$ & 8.95$\times$10$^{-6}$ \\
511&$^{55}$Co, $^{56}$Co, $^{52}$Mn...& 1.91$\times$10$^{-5}$ & 5.94$\times$10$^{-5}$ & 6.95$\times$10$^{-6}$ \\
 1238   &     $^{56}$Co               & 1.47$\times$10$^{-5}$ & 1.10$\times$10$^{-5}$ & 6.00$\times$10$^{-6}$ \\
 1434  & $^{52}$Mn$^g$, $^{52}$Mn$^m$ & 1.04$\times$10$^{-5}$ & 8.32$\times$10$^{-5}$ & 7.33$\times$10$^{-7}$ \\
\enddata
\tablenotetext{a}{Same as Table~4.}
\end{deluxetable}

\end{document}